\documentclass[preprint]{aastex}

\shorttitle{Close binary stars.~VIII}
\shortauthors{Rucinski \& et al.}

\begin{document}

\title{Radial Velocity Studies of Close Binary 
Stars.~VIII\footnote{Based on the data obtained at the David Dunlap 
Observatory, University of Toronto.}}

\author{
Slavek M. Rucinski, 
Christopher C.\ Capobianco\footnote{Current address:
Department of Physics and Astronomy, University of Victoria,
P.O.~Box~3055, STN~CSC, Victoria BC, V8W~3P6, Canada}, 
Wenxian Lu\footnote{Current address:
Dept.\ of Geography and Meteorology and Dept.\ of 
Physics and Astronomy, Valparaiso University,
Valparaiso, IN~46383},
Heide DeBond, J. R. Thomson, Stefan W. Mochnacki,
R.\ Melvin Blake}
\affil{David Dunlap Observatory, University of Toronto\\
P.O.Box 360, Richmond Hill, Ontario, Canada L4C 4Y6}
\email{(rucinski,capobian,lu,debond,jthomson,mochnacki,blake)@astro.utoronto.ca}
\and
\author{Waldemar Og{\l}oza}
\affil{Mt. Suhora Observatory of the Pedagogical University\\
ul.~Podchora\.{z}ych 2, 30-084 Cracow, Poland}
\email{ogloza@ap.krakow.pl}
\and
\author{Greg Stachowski}
\affil{Copernicus Astronomical Center, Bartycka 18, 00--716
Warszawa, Poland}
\email{gss@camk.edu.pl}
\and
\author{P. Rogoziecki}
\affil{Adam Mickiewicz University Observatory, S{\l}oneczna 36, 
60--286 Pozna\'{n}, Poland}
\email{progoz@moon.astro.amu.edu.pl}

\begin{abstract}

Radial-velocity measurements and sine-curve fits to the orbital 
velocity variations are presented for the seventh set of ten close binary 
systems: V410~Aur, V523~Cas, QW~Gem, V921~Her, V2357~Oph,
V1130~Tau, HN~UMa, HX~UMa, HD~93917, NSV~223.
All systems, but three (V523~Cas, HD~93917, NSV~223), were
discovered photometrically by the Hipparcos mission. 
All systems are double-lined (SB2) binaries and all, 
but the detached, very close system V1130~Tau, are contact
binaries. The broadening-function permitted improvement of the orbital
elements for V523~Cas, which was the only system observed
before for radial velocity variations. 
Spectroscopic/visual companions were detected for V410~Aur and HX~UMa.
Several of the studied systems are prime candidates for combined 
light and radial-velocity synthesis solutions.
\end{abstract}

\keywords{ stars: close binaries - stars: eclipsing binaries -- 
stars: variable stars}

\section{INTRODUCTION}
\label{sec1}

This paper is a continuation in a series of papers of
radial-velocity studies of close binary stars 
\citep{ddo1,ddo2,ddo3,ddo4,ddo5,ddo6} and presents
data for the seventh group of ten close binary stars observed
at the David Dunlap Observatory. 
Selection of the targets is quasi-random: At a given time, 
we observe a few dozen close binary systems with periods
shorter than one day, brighter than 11 magnitude and with
declinations $>-20^\circ$; we
publish the results in groups of ten systems as soon as reasonable
orbital elements are obtained from measurements evenly distributed 
in orbital phases. For technical details and conventions,
and for preliminary estimates of errors and uncertainties,
see the interim summary paper \citet[Paper VII]{ddo7}.

This paper is structured in the same way as the previous
papers in that most of the data for the observed binaries are in 
two tables consisting of the radial-velocity measurements 
(Table~\ref{tab1}) and their sine-curve solutions (Table~\ref{tab2}).
Section~\ref{sec2} of the paper contains brief summaries of 
previous studies for individual systems and comments
on the new data. Figures~\ref{fig1} --
\ref{fig3} show the radial velocity data and solutions.
While in the previous papers we showed only some most interesting
broadening functions (BF), Figure~\ref{fig4} in this paper shows
the BF's for all systems; the functions have been 
selected from among the best defined ones around the orbital
phase of 0.25.

The observations reported in this paper have been collected between
March 2000 and April 2002; the ranges of dates for individual 
systems can be found in Table~\ref{tab1}. 
All systems discussed in this paper, except V523~Cas,
have been observed for radial-velocity variations for the first time. 
We have derived the radial velocities in the same way as described 
in previous papers; see Paper~VII for a discussion of the
broadening-function approach used in the derivation of
the radial-velocity orbit parameters:
the amplitudes, $K_i$, the center-of-mass
velocity, $V_0$, and the time-of-eclipse epoch, $T_0$.

The data in Table~\ref{tab2} are organized in the same manner as 
in previous papers. In addition to the parameters of spectroscopic orbits,
the table provides information about the relation between 
the spectroscopically observed epoch of the primary-eclipse T$_0$ 
and the recent photometric determinations in the form of the $O-C$ 
deviations for the number of elapsed periods E. It also contains
our new spectral classifications of the program objects. 

\placetable{tab1}
\placetable{tab2}
\placetable{tab3}

\placefigure{fig1}
\placefigure{fig2}
\placefigure{fig3}
\placefigure{fig4}

\section{RESULTS FOR INDIVIDUAL SYSTEMS}
\label{sec2}

\subsection{V410 Aur}

V410 Aur was discovered by the Hipparcos satellite. The 
somewhat sparsely covered Hipparcos 
light curve is rather typical for a W~UMa-type system; it
has an amplitude of about 0.33 mag. Total eclipses are entirely possible,
given the large semi-amplitudes $K_i$ of our solution, suggesting
an orbital inclination $i \simeq 90$ degrees.
The system is a spectroscopic triple, with the third component having the
relative brightness of $L_3/(L_1+L_2) = 0.26 \pm 0.01$. The
photometric amplitude corrected for the third light is 0.43 mag,
which actually is large and inconsistent with the
the mass-ratio $q_{sp}=0.14$; 
the maximum amplitude for the inclination of
$i=90$ degrees is expected to be around 
0.33 -- 0.38 mag.\ \citep{rci01}.

The individual velocities of the third component did not show any
trace of a ``cross-talk'' (i.e.\ did not correlate with the
binary phase) although the third-star peak in the broadening function was 
strong and always projected into the broadening function
signature of the primary component
(see Figure~\ref{fig4}) making the orbital solution
somewhat poorer than for similar
systems without companions (see Figure~\ref{fig1}). 
The radial velocity observations of the
third component are given in Table~\ref{tab3}.
Our observations were made in two groups,
around HJD~2,451,870 and HJD~2,452,280. The average velocities of the
third component for these groups were 
$V_3 = 48.38 \pm 0.30$ km~s$^{-1}$ and
$43.90 \pm 0.43$ km~s$^{-1}$; both differ 
significantly from the mean velocity
of the close binary, $V_0 = 36.94 \pm 3.10$ km~s$^{-1}$. Unfortunately,
we do not have enough observations and they are not accurate enough
to check if the difference in the average $V_3$ is reflected in a
change of the binary center-of-mass velocity $V_0$; our solution
for the binary uses one constant value for $V_0$.

The orbital ephemeris based on the Hipparcos data shows a large
$O-C$ deviation of 0.10 days. We arbitrarily reduced the $O-C$ by shifting
the time of the primary eclipse by that amount under 
an assumption that V410~Aur is a contact binary of the A type 
(i.e.\ that the hotter component is the more massive one).

The third body in the system was probably the reason for a large error
in the Hipparcos parallax, $4.77 \pm 5.39$ mas, so that we have no
distance estimate for the system. There is a discrepancy between
the $V_{max}$ estimated from the Hipparcos $H_P=9.98$ and from
the Tycho--2 \citep{tyc2} photometry, $V_{max}=10.18$ (transformed
from $V_T$). Our estimate of
the spectral type, G0/2V, is consistent with the average
$(B-V)=0.56$ from Tycho--2. 

\subsection{V523 Cas}

V523 Cas is the only system in this group which was observed
spectroscopically before our observations \citep{mil85}. The system is
an interesting one because -- with its very short
orbital period of 0.237 days, right next to CC~Com at 0.221 days -- it
is very close to the still-unexplained abrupt cutoff in the
orbital period distribution of contact binaries.
Several photometric studies of V523~Cas
have been recently rediscussed by \citet{lis00} where all previous
references are given. The photometric solutions encountered 
the common problem of a featureless light curve 
poorly constraining the geometrical
parameters, with the crucial parameter of the mass ratio
taking a large range of values, with the photometric ones usually
above the value determined by Milone et al.\ of
$q_{sp} = 0.42 \pm 0.02$.

Our radial velocity results are well defined (Figure~\ref{fig1}). 
The broadening function (Figure~\ref{fig4})
very clearly shows the more compact and slightly hotter
(or having a higher surface brightness) less-massive component.
The disparity of the component shapes is quite striking 
in the broadening function; it seems as if
the secondary were a detached component (but still quite large) 
in a semi-detached binary. Our spectroscopic
mass ratio, $q_{sp} = 0.516 \pm 0.008$, 
is very close to the photometric value determined most
recently by \citet{lis00}, $q_{ph} = 0.53 \pm 0.02$, hopefully
ending the long dispute about the $q_{sp}$ versus $q_{ph}$ 
disparity \citep{mac86}. 

The ephemeris used to phase the observations
was based on two sources: The period, as
given in Table~\ref{tab2}, is from \citet{nel01} while the
initial epoch, $T_0=2,451,822.4267$, was taken from \citet{pri01}.

V523~Cas does not have a trigonometric parallax and was apparently
too faint to be included in the Tycho--2 project. Thus, we 
refrain from a full discussion of its parameters, noting only that
our spectral type, K4V, confirms previous classifications and
agrees with $(B-V)=1.07$ measured by \citet{bra81}; in that work
$V_{max}=10.59$. The (extrapolated) calibration of \citet[RD97]{RD97}
gives $M_V=6.15$ so that V523~Cas is one of the faintest known
contact binaries. This gives a rather small distance to the
binary, about 77 pc.

\subsection{QW Gem}

This contact system has been discovered by Hipparcos. It is
somewhat faint for our telescope, so that the individual 
broadening functions are noisy, but the orbital solution
is well defined (Figure~\ref{fig1}). 
This contact system is of the W-type when 
the Hipparcos ephemeris is used; this ephemeris gives a very small
$O-C$ deviations. The light curve has a moderately large amplitude 
of 0.41 mag.

The maximum light brightness,
$V_{max}=10.3 \pm 0.1$, is estimated from the transformed $H_P$ data and
the Tycho--2 average data; this estimate is very approximate. The 
Hipparcos parallax has a large error, $4.09 \pm 5.42$, which
may indicate some undiscovered companions or just reflects
the relative faintness of the star. Our spectral type,
F8V, is consistent with the Tycho--2 average $(B-V)=0.48$.
$M_V=3.36$ determined from the very uncertain parallax
is in agreement with $M_V(cal)=3.55$ derived using the RD97
calibration.

\subsection{V921 Her}

V921 Her is another Hipparcos discovery. This contact system is 
of the A-type with the more massive component eclipsed during
the deeper minimum. The Hipparcos light curve is well defined,
with an amplitude of 0.36 mag.
The system is interesting in that it has a relatively long period
of 0.877 days which is rather infrequent among contact binaries.
The color index estimated from the Tycho--2 data, $(B-V)=0.27$, points
at a spectral type slightly later than our new classification, 
A7IV (the SIMBAD type is an even earlier A5), 
but there may be some interstellar reddening. 
The Hipparcos parallax, $2.39 \pm 0.97$ mas
and $V_{max}=9.37$, with an assumption of
no reddening, give $M_V=1.3 \pm 0.9$,
which is consistent with the extrapolated
RD97 calibration, $M_V(cal)=1.1$. This is one
of the most luminous contact systems with $P < 1$ day. 

The components
are only barely resolved in the broadening function which may indicate
a low orbital inclination (a small $\sin i$).  
The orbit is moderately well defined (Figure~\ref{fig1}), 
partly due to the early spectral type and
dearth of lines; both factors tend to increase the velocity errors. An 
interesting feature is a relatively large (in the absolute sense)
mean systemic velocity of $V_0 = -79$ km~s$^{-1}$; the Hipparcos
tangential velocities are small so that the large radial velocity
is the main contributor to the spatial velocity of 81 km~s$^{-1}$.

\subsection{V2357 Oph}

This variable was discovered by Hipparcos. It was classified
as a pulsating star with a period of 0.208 day which is
two times shorter than the actual orbital period. The Hipparcos
ephemeris was for the maximum light. Assuming that the
Hipparcos initial epoch corresponded to the minimum just before
the maximum used in the ephemeris ($T_0 = 2,448,500.0451$), the
resulting $O-C$ is quite small, see Table~\ref{tab2}.

The orbit is only moderately well defined
(Figure~\ref{fig2}), partly because the system
is a bit faint for our instrumentation, and partly because of
the small separation of the components in the broadening
functions. The small radial velocity semi-amplitudes $K_i$ and the
small photometric amplitude of 0.12 mag.\ are all consistent
with the low orbital inclination.  

The Hipparcos parallax, $5.37 \pm 2.01$ mas and $V_{max}=10.43$
lead to $M_V=4.08 \pm 0.82$, while $M_V(cal)=4.23$ (RD97) for the
average $(B-V)=0.80$, as indicated by the Tycho--2 data. This
color index is approximately consistent with our spectral type,
G5V.

\subsection{V1130 Tau}

This bright ($V_{max} = 6.56$) system was discovered by Hipparcos.
It is the only non-contact system in this group of ten
binaries. The broadening functions are exceptionally well defined 
and show (Figure~\ref{fig4}) two large, but detached components
of similar size and brightness. The mass ratio resulting from
our orbit (Figure~\ref{fig2}) is 
$q=0.919 \pm 0.008$. The photometric variations observed by Hipparcos
define a very nice light curve with rounded maxima, 
and an amplitude of of 0.38 mag. The distinction of
which component is the hotter one and thus which of the minima should
be called the primary minimum is not well established from
the Hipparcos light curve. The broadening functions show the
fainter component eclipsed during the Hipparcos' primary minimum,
which may indicate that the identification of the eclipses is 
actually incorrect. 

The system begs to be analyzed in detail as it is one of the shortest
period detached systems with the spectral type F3V. This
spectral type is
in agreement with the average $(B-V)=0.34$, as derived from the
Tycho--2 data. The Hipparcos parallax is relatively large (in
fact the largest in this group of binaries) and 
well determined, $15.35 \pm 0.78$ mas, leading to $M_V=2.49 \pm 0.12$.
The contact-binary calibration of RD97 obviously does not apply in this
case; we can only note that the combined brightness corresponds
to some 0.6 mag.\ above the Main Sequence, which simply reflects the
detached binary nature of two similar stars.

We note that $(M_1+M_2) \sin^3i = 2.41 \pm 0.03 \,M_\odot$
for V1130~Tau is relatively large so that the orbital
inclination must be close to 90 degrees.

\subsection{HN UMa}

Another Hipparcos discovery, HN~UMa, is a rather commonly-looking 
contact system seen at moderately low orbital inclination because
both the photometric variations (0.12 mag.) as well as the radial
velocity semi-amplitudes, $K_i$, are rather small (Figure~\ref{fig2}). 

The spectral type of F8V is somewhat early for the average 
$(B-V)=0.46$ derived from the Tycho--2 data. The Hipparcos parallax,
$5.81 \pm 1.39$ mas, with the adopted $V_{max}=9.80$, give
$M_V=3.62 \pm 0.53$, while $M_V(cal)=3.36$ (RD97). 
The proper motion of the system (Tycho--2 data) is comparatively 
large, $\mu_\alpha\cos(\delta)=+54.0 \pm 1.9$ mas and 
$\mu_\delta=-36.9 \pm 1.8$ mas, which together
with $V_0=-37.1$ km~s$^{-1}$ results in a relatively large
spatial velocity of 65.0 km~s$^{-1}$.

\subsection{HX UMa}

The variability of HX~UMa was discovered by Hipparcos. The light
curve is of a typical W~UMa-type system with an amplitude of
0.17 mag. Apparently, the ephemeris of Hipparcos predicts the
photometric minima quite well with a very small
$O-C$ shift; the deeper one corresponds to the eclipse of the
more massive star.

A close companion at the distance of 0.63 arcsec was
discovered by Hipparcos. The re-determination by \citet{fab00}
gave an average magnitude difference between the components 
of 3.00 mag.\ in the $H_P$ system; 
our determination of a weak third-light peak in the
broadening function gives
$L_3/(L_1+L_2)=0.049 \pm 0.004$ or $\Delta m = 3.27 \pm 0.10$
at the bandpass of our spectroscopy at 5184 \AA.
The signature of the third body was just barely detectable 
(Figure~\ref{fig4}) and, if not for the persistence 
through the orbital phases of the
binary, it could be easily taken for noise in the broadening
function. The radial velocity data for the third component
are given in Table~\ref{tab3}. The average velocity,
$V_3 = -23.59 \pm 0.77$ km~s$^{-1}$, is slightly
different from the center of mass velocity of the binary,
$V_0 = -19.88 \pm 0.39$ km~s$^{-1}$, 
but the difference may  be of no
significance as we cannot exclude any systematic 
bias in the measurements of $V_3$ on top of the much
stronger feature of the close binary system.
We do not see any ``cross-talk'' in the velocity
determinations in the sense that $V_3$
does not depend on the orbital phase of the close binary
(see Paper~VII for a discussion of cases when such
a dependence is observed).
Our observations were made in two groups, centered
on HJD 2,451,697.4 and 2,451,986.9 (see Table~\ref{tab3}). 
We do not see any
significant change in $V_3$ between these epochs, with
the average values of $-22.95 \pm 1.12$ and $-24.56 \pm 1.00$
km~s$^{-1}$.

The spectral type, F8V, is consistent with $(B-V)=0.44$
derived from the Tycho--2 data. The Hipparcos parallax is
poorly determined, $6.68 \pm 3.01$, possibly reflecting the
visual-binary nature of the system, so it cannot be
meaningfully compared with the RD97 calibration which
predicts $M_V(cal)=3.32$.

It was noted by \citet{BL97} that HX~UMa is not 
physically related to the
nearby variable star DP~UMa (HD~104513).

\subsection{HD 93917}

The W~UMa-type variability of this star has been recently 
discovered by \citet{la01}; we used the timing data from that
work for the values of $T_0$ and $P$. The amplitude of light variations
is 0.34 mag.\ so it is rather curious that the variability 
of this bright ($V_{max} = 9.02$) 
star went undetected for such a long time. There
are more curious circumstances around this star: It is one
of very few HD stars not observed by the main 
Hipparcos project (but it was observed by Tycho--2). Also, there
existed a large discrepancy between the $(B-V) = 0.55$ color
index, as given in the SIMBAD database and confirmed by 
Tycho--2, and the SIMBAD spectral type of K0. We found that the spectral
type is F9.5V which agrees very well with the color index.
The predicted $M_V(cal)=3.35$ (RD97) cannot be checked because
the Tycho parallax is poor, $14.9 \pm 11.4$ mas.

Our observations define a very good spectroscopic orbit
without any peculiarities (Figure~\ref{fig3}). 
The system belongs to the W-type
of contact binaries, i.e.\ has a hotter less-massive component.
There is absolutely nothing peculiar in the system except for the
very fact that, being so bright, its variability remained undetected 
for such a long time. We note that \citet{ruc02} (Sec.~8 \& 12) 
estimated that even among brights stars within the interval
$7.5 < V < 8.5$, perhaps as many as 45 
contact systems still remain to be discovered.

\subsection{NSV 223}

The variability of NSV~223 was known since the discovery
by \citet{str56}. Recently, \citet{vv00} obtained a
light curve showing a total, long-duration
eclipse in the secondary minimum, which
suggests that the mass ratio is small. Our radial velocity data
(Figure~\ref{fig3}) fully confirm these predictions: 
The contact system is indeed of the
A-type and the mass ratio is small, $q=0.136 \pm 0.011$. 
The large semi-amplitudes $K_i$ suggest that the orbital
inclination is close to 90 degrees. In spite of those properties
promising a very good combined solution, the system is poorly known.
What is known is $V_{max}=10.86$ from \citet{vv00}; also
our spectral type is F7V. The ephemeris given by the
discoverers holds quite well (see Table~\ref{tab2}).

\section{SUMMARY}

This paper presents radial-velocity data and orbital solutions 
for the seventh group of ten close binary systems 
observed at the David Dunlap Observatory. 
In this series, only V523~Cas has a history of 
previous radial-velocity studies. Seven system were 
discovered photometrically by the Hipparcos satellite mission while
two binary systems, HD~93917 and NSV~223
have been identified as contact systems only very recently.

All systems of this group, with the exception of the
detached binary V1130~Tau, are contact binaries
of the W~UMa type. All systems are double-lined (SB2) 
binaries with visible spectral lines of both components.
V1130~Tau is interesting one because it is a very close 
yet detached consisting of two almost identical F3V stars 
on a tight orbit with $P=0.799$ days. In contrast,
the early-type (A7) contact binary V921~Her  
has a relatively long period of $P=0.877$
days. A bright (9 mag.), recently discovered contact
system HD~93917 is rather inconspicuous in its properties
except for the fact that it remained
undiscovered for such a long time; this confirms the
prediction that many close binary systems still remain to be
discovered among bright stars. 
V410~Aur and HX~UMa have fainter companions detected 
spectroscopically.

 
\acknowledgements

Support from the Natural Sciences and Engineering Council of Canada
to SMR and SWM is acknowledged with gratitude. 

The research has made use of the SIMBAD database, operated at the CDS, 
Strasbourg, France and accessible through the Canadian 
Astronomy Data Centre, which is operated by the Herzberg Institute of 
Astrophysics, National Research Council of Canada.

\clearpage

\noindent
Captions to figures:

\bigskip

\figcaption[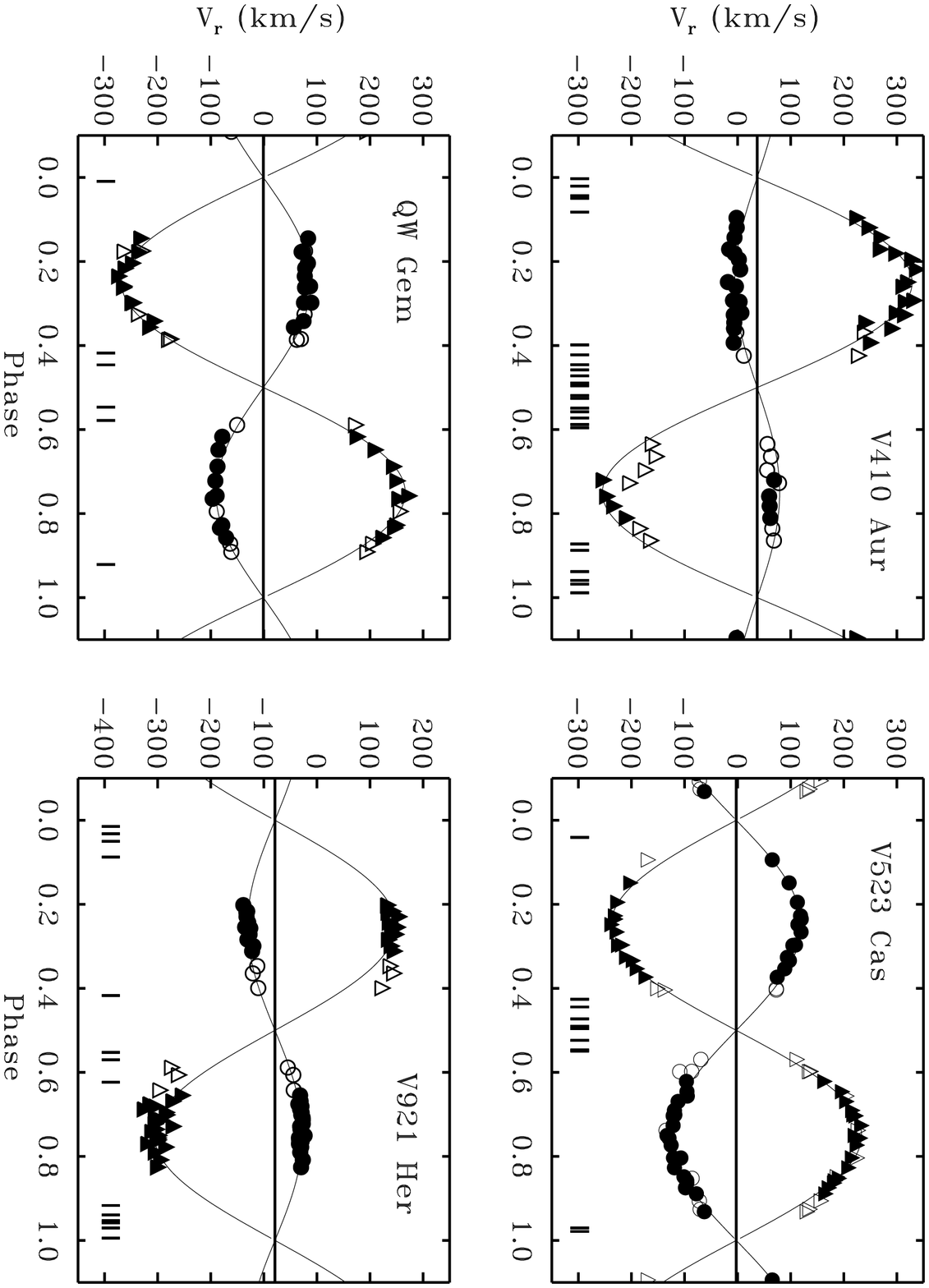] {\label{fig1}
Radial velocities of the systems V410~Aur, V523~Cas, QW~Gem and
V921~Her are plotted in individual panels versus the orbital phases. 
The lines give the respective circular-orbit (sine-curve) 
fits to the radial velocities. All four systems are
contact binaries. V410~Aur and V921~Her are A-type contact systems, 
while V523~Cas and QW~Gem are W-type contact system. V410~Aur
has a fainter spectroscopic companion whose presence lowered
the quality of the measurements and of the solution.
The circles and triangles in this and the next two figures
correspond to components with velocities $V_1$ and $V_2$,
as listed in Table~\ref{tab1}, respectively. 
The component eclipsed at the minimum corresponding to 
$T_0$ (as given in Table~\ref{tab2}) is the one which shows
negative velocities for the phase interval $0.0 - 0.5$.
The open symbols indicate observations contributing 
half-weight data in the solutions. Short marks in the lower 
parts of the panels show phases of available 
observations which were not used in the solutions because of the 
blending of lines. All panels have the same vertical 
range, $-350$ to $+350$ km~s$^{-1}$, but the display is vertically
shifted for V921~Her.
}

\figcaption[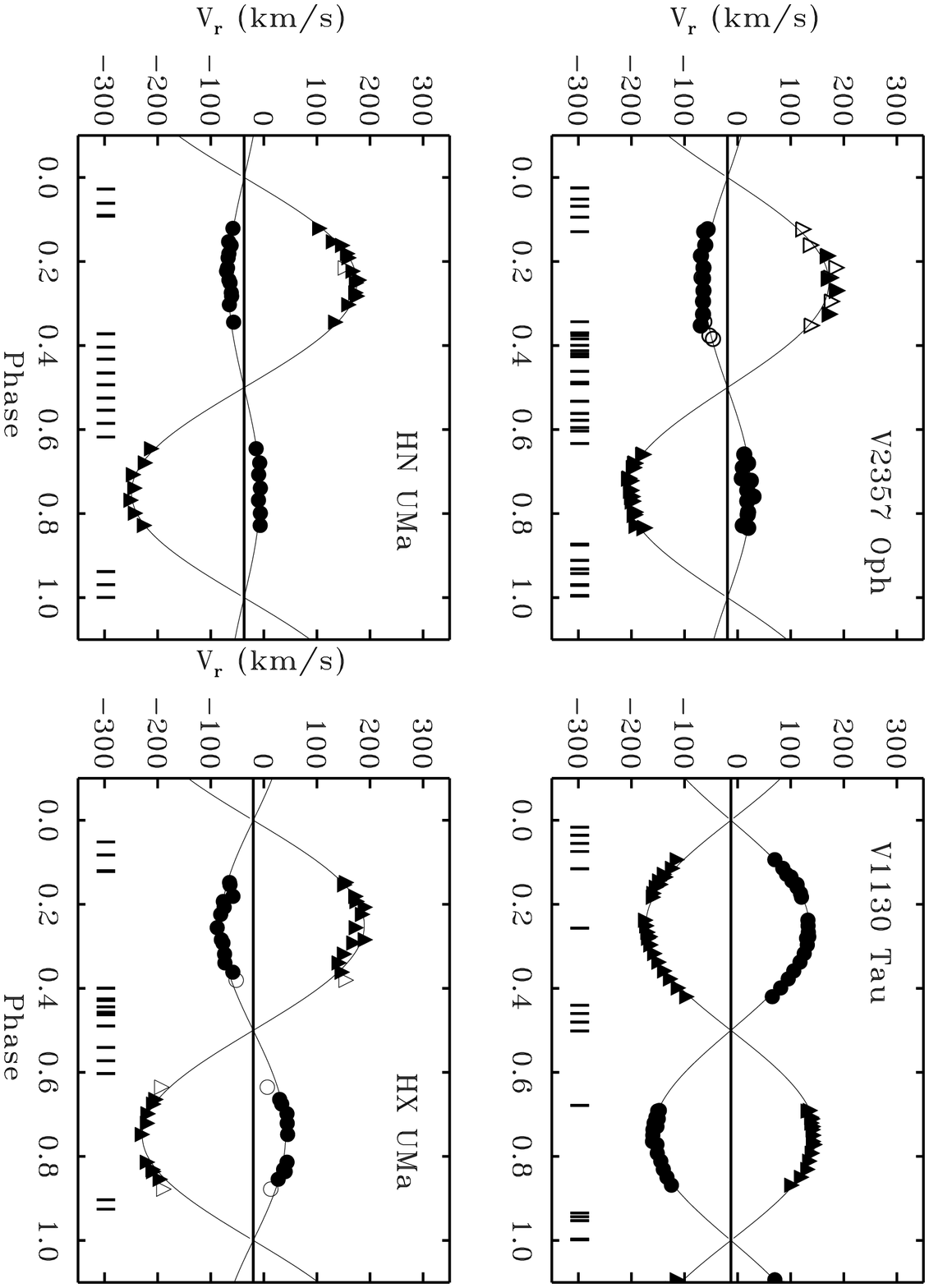] {\label{fig2}
Same as for Figure~1, but with the radial velocity orbits
for the systems V2357~Oph, V1130~Tau, HN~UMa and HX~UMa.
All systems, except V1130~Tau, which is a close detached
system, are contact systems of the A-type. HX~UMa has a faint
companion.
}

\figcaption[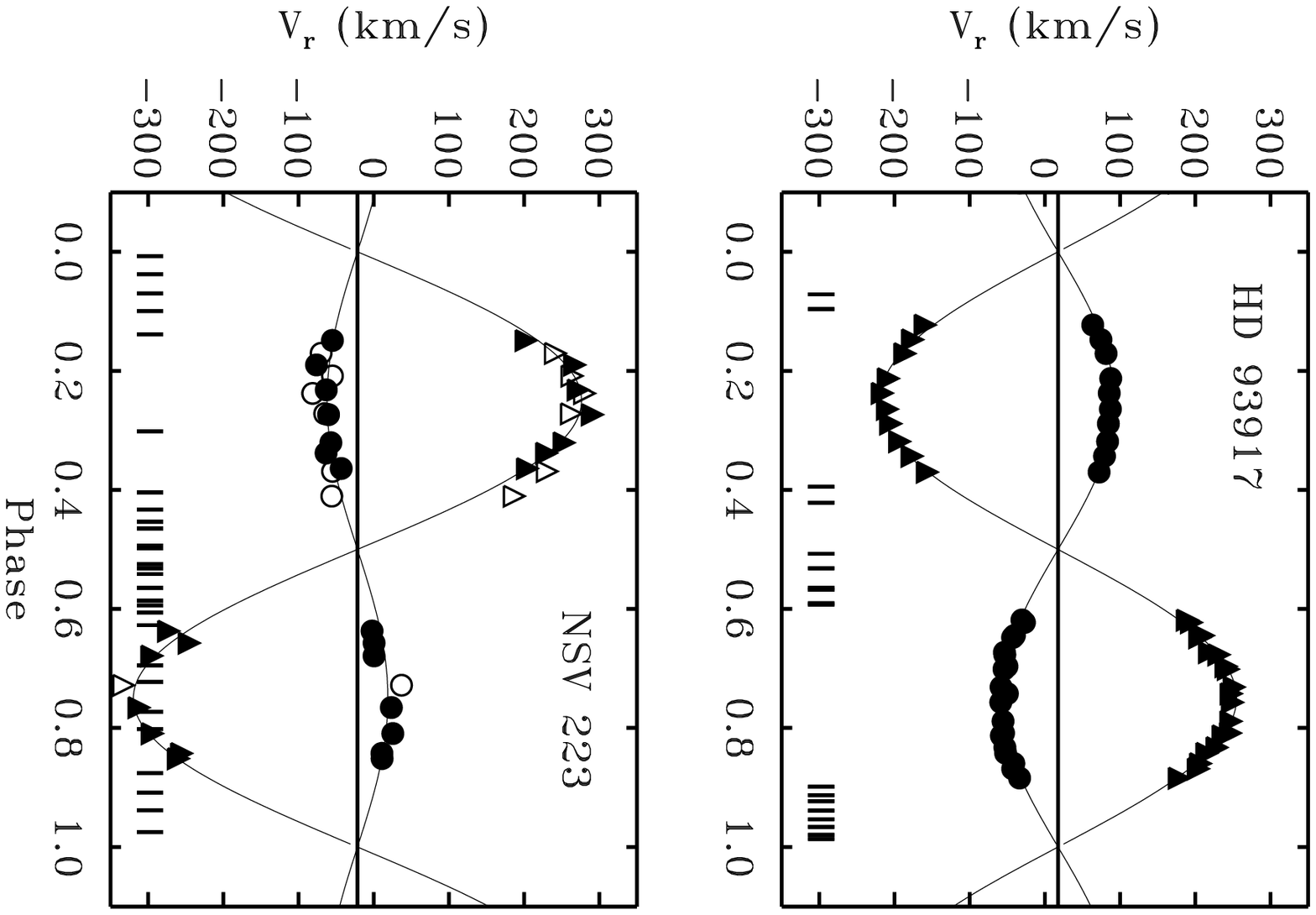] {\label{fig3}
Same as for Figure~1, for the systems HD~93917 and NSV~223.  
The former is a contact binary 
of the W-type, while the latter is of the A-type.
}

\figcaption[fig4_60w.ps] {\label{fig4}
The broadening functions (BF's) for all binary systems of this
group, all for orbital phases around 0.25, as in similar figures
in the previous papers. V410~Aur has a well defined 
slowly rotating companion. A tiny third-body feature in the
broadening function of HX~UMa could easily be taken for a small error 
fluctuation (as in poorly determined cases of fainter systems,
such as NSV~223), if not for its persistence throughout 
all the phases. 
}

\clearpage                    

\begin{table}                 
\dummytable \label{tab1}      
\end{table}
 
\begin{table}                 
\dummytable \label{tab2}      
\end{table}

\begin{table}                 
\dummytable \label{tab3}      
\end{table}

\end{document}


\begin{deluxetable}{lccrrrccc}

\tabletypesize{\scriptsize}

\pagestyle{empty}
\tablecolumns{9}

\tablewidth{0pt}
\tablenum{2}
\tablecaption{Spectroscopic orbital elements}
\tablehead{
   \colhead{Name} &                
   \colhead{Type} &                
   \colhead{Other names} &         
   \colhead{$V_0$} &               
   \colhead{K$_1$} &               
   \colhead{$\epsilon_1$} &        
   \colhead{T$_0$ -- 2,400,000} &  
   \colhead{P (days)} &            
   \colhead{$q$}          \\       
   \colhead{}     &                
   \colhead{Sp.~type}    &         
   \colhead{}      &               
   \colhead{} &                    
   \colhead{K$_2$} &               
   \colhead{$\epsilon_2$} &        
   \colhead{(O--C)(d)~[E]} &         
   \colhead{$(M_1+M_2) \sin^3i$} &    
   \colhead{}                      
}
\startdata

V410 Aur   & EW/A  & HD~280332 & $+36.94$(3.10) & 42.14(3.73) &
       11.53 & 52,061.1068(22) & 0.366340 & 0.144(13) \\
             & G0/2V & HIP~23337 &          & 291.68(4.50) &
       22.19 & $+0.1060$~[9720] & 1.42(10) &          \\[1mm]

V523 Cas   & EW/W  &           & $-2.54$(0.90)    & 121.64(1.14) &
       7.82  & 52,104.0295(4)  & 0.233693 & 0.516(7)  \\
             & K4V &           &          & 235.95(1.41) &
      10.37  & $+0.0027$~[1205] & 1.110(24) &         \\[1mm]

QW Gem     & EW/W  & HD~264672 & $-0.90$(1.39)   &  89.17(1.84) &
       7.79  & 51,961.2357(11) & 0.358127 & 0.334(9) \\
             & F8V & HIP~32845 &          & 267.30(2.52) &
      12.14  & $-0.0096$~[9664] & 1.685(62) &         \\[1mm]

V921 Her   & EW/A  & HD~152172 & $-79.04$(1.05)  & 51.45(0.89) &
       4.06  & 51,979.8304(62) & 0.877366 & 0.226(5) \\
             & A7IV: & HIP~82344 &        & 227.17(1.97) &
      17.37  & $+0.0718$~[3966] & 1.971(61) &         \\[1mm]

V2357 Oph   & EW/A  &          & $-19.12$(1.64)  & 44.12(1.63) &
       7.37  & 52,050.2364(22) & 0.415568\tablenotemark{a} & 0.231(10) \\
             & G5V & HIP~82967 &          & 190.93(2.90) &
       9.93  & $-0.0061$~[8543] & 0.560(32) &         \\[1mm]

V1130 Tau   & EA    & HD~24133 & $-12.74$(0.46)  & 147.21(0.63) &
       3.39  & 52,097.6133(11) & 0.798871 & 0.919(7) \\
             & F3V & HIP~17988 &             & 160.11(0.74) &
       4.26  & $-0.0218$~[4503] & 2.408(32) &         \\[1mm]

HN UMa      & EW/A  & BD$+38^\circ 2220$ & $-37.11$(0.63) & 29.65(0.76) &
       3.02  & 52,049.9590(8)  & 0.382608 & 0.140(4) \\
             & F8V & HIP~55030 &             & 212.22(0.96) &
       4.42  & $+0.0439$~[9278] & 0.562(12) &         \\[1mm]

HX UMa      & EW/A  & HD~104425 & $-19.88$(1.11) & 60.80(1.45) &
       6.16  & 51,835.0458(13) & 0.379156 & 0.291(9) \\
             & F4V & HIP~58648 &             & 209.18(2.07) &
       9.53  & $-0.0032$~[8795] & 0.775(30) &         \\[1mm]

HD~93917    & EW/W  & BD$-1^\circ 2452$\tablenotemark{b} & $+17.47$(0.79) & 74.33(0.91) &
       4.12  & 52,374.6151(8) &  0.443420 & 0.313(5) \\
             & F9.5 &            &           & 237.30(1.32) &
       6.12  & $+0.0095$~[810]   & 1.394(30) &         \\[1mm]

NSV~223     & EW/A  & BD$+20^\circ 75$ & $-21.66$(2.15) & 40.39(2.64) &
       9.02  & 50,748.8907(18)    & 0.366128  & 0.136(10) \\ 
             & F7V  &            &           & 297.98(4.18) &
      14.43  & $-0.0053$~[2941]  & 1.47(9)   &         \\[1mm]
 
\tablenotetext{a}{V2357 Oph: The period is twice as long as
in the Hipparcos catalogue whare the star was classified as
pulsating one. The $T_0$ has been adjusted
from the maximum to the minimum of light, as described in the text.}
\tablenotetext{b}{HD~93917: The variability discovery paper gives
BD$-1^\circ 2448$ while SIMBAD gives BD$-1^\circ 2452$.}

\tablecomments{The spectral types given in column two
are all new and relate to the combined spectral
type of all components in a system. The convention of naming the
binary components in this table is that the more massive
star is marked by the subscript ``1'', so that 
the mass ratio is defined to be always $q \le 1$. Figures 1 -- 3
should help identify which component is eclipsed at the primary
minimum. The standard errors of the circular 
solutions in the table are expressed in units of last decimal places 
quoted; they are given in parantheses after each value. 
The center-of-mass velocities ($V_0$), the 
velocity amplitudes (K$_i$) and the standard unit-weight 
errors of the solutions ($\epsilon$) are all expressed 
in km~s$^{-1}$. The spectroscopically determined 
moments of primary minima are given by $T_0$; the corresponding
$(O-C)$ deviations (in days) have been calculated from the most 
recent available ephemerides, as given in the text,
using the assumed periods and the number of epochs given by [E].
The values of $(M_1+M_2)\sin^3i$ are in the solar mass units.
}
\enddata
\end{deluxetable}